\documentstyle[epsfig,natbib209]{mn}
\newcommand{\be}{\begin{equation}}
\newcommand{\ee}{\end{equation}}
\newcommand{\ba}{\begin{eqnarray}}
\newcommand{\ea}{\end{eqnarray}}

\title{Pseudo-Newtonian Models for the Equilibrium Structures 
of Rotating Relativistic Stars}

\author{Jinho Kim, Hee Il Kim, and Hyung Mok Lee\\
Department of Physics and Astronomy, FPRD, Seoul National 
University, Seoul
  151-747, Korea \\
{\it email}: jinho@astro.snu.ac.kr, khi@astro.snu.ac.kr and 
hmlee@astro.snu.ac.kr}
\begin{document}
\maketitle
\begin{abstract}
 We obtain equilibrium solutions for rotating compact stars including 
 the special relativistic effects. The gravity is assumed to be 
Newtonian, but we used the active mass density, which takes into account
all the energies such as motions of the fluids, internal energy, pressure
energy in addition to the rest mass energy, in computing the gravitational
potential using Poisson's equation. 
 Such a treatment could be applicable to the neutron stars with 
 relativistic motions or relativistic equation of state.
 We applied the Hachisu's self-consistent field (SCF) method to 
 find spheroidal as well as toroidal sequences of equilibrium solutions.
 Our solutions show better agreement than Newtonian relativistic 
 hydrodynamic approach that does not take into account the active mass,
 with general relativistic solutions. The physical quantities such as the
peak density, equatorial radii of our solutions agree with general
relativistic ones within 5\%.Therefore our approach can be a
simple alternative to the fully relativistic one when large
number of model calculations are necessary as it requires much
less computational resources.   
\end{abstract}


\section{Introduction}
Finding an equilibrium solution is a starting point for the studies of the 
dynamical evolution of any objects.
The sequence of equilibrium models as a function of particular parameter could 
also give us some guess
and insight for the real evolutionary dynamics when direct numerical simulations are difficult.
It is rather trivial to get solutions for non-rotating spherical stars
because it only requires the integration of a single ordinary differential equation 
with proper boundary conditions if the equation of state is barotropic. 
For rotating stars having non-spherical shapes, however, 
solving the elliptic equation for the problem is not so simple 
because the location of boundary of 
the star is not predetermined on which the boundary condition should 
be imposed. Since \cite{jam65} first found the solution of 
rotating stars by directly integrating the elliptic equation, 
many people have tried to develop more
efficient and accurate ways.
Self-consistent field (SCF) method \citep{ost68}, which is one of 
the possible approaches, uses integral
representation rather than solving the differential equations directly.
Based on SCF approach, \cite{hac86a,hac86b} developed a very successful 
method which covers almost all possible configurations of rotating 
stars with barotropic equation of state and Newtonian gravity. 
Soon after it was devised, this method was adopted to generate general
relativistic initial data of rotating stars (\citealt{kom89a,kom89b}, 
KEH hereafter). It was also used for the study of supramassive stars 
which cannot be treated with Newtonian gravity 
\citep{coo92,coo94a,coo94b} and for the study of rotating stars with 
realistic equation of state \citep{coo94a,coo94b,ste95}.
Now, SCF is one of the most popular methods to find equilibrium 
solutions for wide ranges of problems other than single rotating 
stars e.g., binary stars \citep{bau98}, magnetized stars \citep{kiu08}.

Relativistic motions of self-gravitating objects are possible in 
various astrophysical circumstances. During the core collapse of 
proto-neutron stars, the fluid velocity can reach 
$\sim 0.2 c$ where $c$ is the speed of light.
Recently found millisecond pulsar XTE J1789-285 has rotation frequency 
of 1122 Hz \citep{kaa07}, which corresponds to the rotation speed 
$\sim 0.2-0.4 c$ at the stellar surface. Many phenomenological 
models for neutron stars require relativistic equation of state
whose pressure and internal energy density can be comparable to 
the rest mass energy density.
All these ingredients should be well implemented in general 
relativistic hydrodynamics. It is now possible to study these 
objects in three-dimensional numerical
simulations by taking into account the dynamics of
spacetime [see for a review, \cite{fon08}].
However, it is very difficult to simulate realistic details of neutron 
star physics general relativistically
because it requires too much computational resource.
Therefore, Newtonian gravity is still widely used in cases the 
gravitational fields are not so strong \citep{toh90,ou06,pet08}. 
Sometimes Newtonian approach has been adopted even for
the problems in which relativistic motions or relativistic 
equation of states are involved because of its simplicity. 
This approach might be justified if it does not produce significantly 
different results from the fully relativistic calculations.
However, it is better to find a new approach which can account 
for the relativistic properties in a generic way.

The purpose of this paper is to implement the special relativistic 
effects properly in finding the equilibrium solutions for rapidly 
rotating compact stars under the Newtonian gravity. Our approach
is different from previous studies in the sense that we use the
active mass which takes into account the contributions from
various energies to the gravitational potential, in addition
to the rest mass energy. Our framework is based on the Hachisu's
SCF method. The implememtation
of active mass should provide an improved agreement with the general 
relativistic solutions since the concept of active mass essentially
comes from the general relativity. 
The rotating neutron stars and strange stars would be the 
best examples to which our method can be applied.

This paper is organized as follows. In section 2, we describe our 
approach taken to find a hydrostatic equilibrium solution of 
rotating stars. The details of numerical scheme to find the solutions are 
given in section 3. We compare our solutions with those obtained by Newtonian 
and general relativistic approaches in section 4.
The properties of our solutions for the rotating stars are presented and 
discussed in \S 5. The final section provides summary and concluding 
remarks.

\section{Formulation}
The general relativity is the only way to incorporate the special 
relativity with gravity. The dynamics of the gravity (or spacetime) 
can be analyzed by solving the Einstein equations.
At the same time, the equations of motion for matter are given by the 
conservation of the energy-momentum tensor which is the source of 
the gravity. All these are highly coupled and nonlinear equations 
so that the state of the arts numerical techniques and massive 
computational resource are required. If the gravity is weak, however, 
one can take the so called ``weak field approximation'' by 
linearizing the Einstein equations and solving only the matter 
dynamics in flat spacetime. Although it is now much simpler and 
sometimes is possible to treat analytically, we still have
6 unknowns for the spacetime (after gauge fixing) in general and 
have 4 unknowns for axisymmetric rotating stars of our concern. 
But this is too many for our purpose.
Therefore, just like the Newtonian approach, we take only one dynamical 
variable (gravitational potential) for the spacetime by assuming the 
following metric,
\be
ds^2 = -(1+2\Phi)dt^2 + (1+2\Phi)^{-1}\delta_{ij}dx^{i}dx^{j},
\ee
where $\Phi$ is the Newtonian gravitational potential.
In the weak field approximation, the assumption results in 
$T_{ij},~~T_{i0} \ll T_{00}$. This condition can be violated by 
the relativistic problems of our concern.
Since we try to develop an effective method, however, we shall not 
stick to this discrepancy. Instead, we take into account the possible 
relativistic circumstances of $T_{ij},~~T_{i0} \sim T_{00}$
by introducing ``{\it{active mass density}}'', $\rho_{\rm{active}}$ 
which was first suggested by \cite{tol}.
The opposite concept, ``{\it{passive mass density}}'' is just the 
inertial rest mass density which appears as a source term in the 
Poisson's equation. However, it is clear from the general relativity 
and the mass-energy equivalence that all kinds of energies can 
contribute to gravitational potential that influences spacetime geometry.
The corresponding active mass density given by \cite{tol} has the form
\be
\rho_{\rm{active}} = T - 2T^{0}_{0} = T^{i}_{i} - T^{0}_{0} ,
\ee
where $T$ is the trace of the energy-momentum tensor.
Accordingly, we modify the Poisson's equation to have 
the $\rho_{\rm{active}}$ in the source term,
\be
\nabla^2 \Phi = 4\pi \textrm{G}\rho_{{\rm{active}}} .
\ee

To describe the rotating star we assume that the energy-momentum 
tensor of matter is that of perfect fluid as given by
\be
T^{\mu\nu} = \rho_0 H u^\mu u^\nu + P g^{\mu\nu} ,
\ee
where $\rho_0$ is the rest mass density, which is proportional to
the number density of baryon of the fluid,  $P$ is the pressure,
$u^\mu$ is the four velocity of a fluid element with respect to the
Eulerian observer and $g_{\mu\nu}$ is the spacetime metric assumed 
in eq. (1), and
$H$ is the specific enthalpy which is defined as
\be
H = 1+\epsilon + \frac{P}{\rho_0} ,
\ee
with $\epsilon$ being the specific internal energy.
The equation of state is assumed to be barotropic, i.e.,
$P = P( \rho_0 )$. Furthermore, for convenience, we only consider 
simple Polytropic equation of state,
\be
P = \kappa \rho_{0}^{1 + \frac{1}{N}} ,
\ee
where $\kappa$ and $N$ are polytropic coefficient and index, respectively.

From the general relativistic version of hydrostatic equation found 
by KEH, we can easily get the hydrostatic equation for our metric 
as follows,
\ba
\frac{1}{\rho_0 H}\nabla P &+& \nabla\ln\left(1+2\Phi\right) \nonumber \\
&-&\frac{v}{1-v^2}\nabla v +\frac{v^2}{1-v^2}\frac{\nabla\Omega}{\Omega} = 0 ,
\ea
where $\Omega$ is the rotational angular speed, 
and $v=\Omega r \sin\theta/\left(1+2\Phi\right) $.
The fluid four-velocity $u^\mu$ due to the rotation is thus given by
\be
u^\mu = \frac{1}{\sqrt{\left(1+2\Phi \right)\left(1-v^2\right)}} (1,0,0,\Omega) .
\ee
The axial symmetry is assumed and the spherical polar coordinates 
$(r,\theta, \phi)$ are used. If the last term of the left hand side of 
equation (7) is a function of $\Omega$, the hydrostatic equation is 
integrable. Usually, the $\Omega$-dependence is assumed to have the 
following form (KEH),
\be
F\left(\Omega\right)=A^2\left(\Omega_C-\Omega\right)=\\
\frac{\Omega r^2 \sin^2\theta}{\left(1+2\Phi\right)^2-\Omega^2 r^2 \sin^2\theta} ,
\ee
where $\Omega_C$ is the angular speed at the rotation axis, 
and $A$ is a constant which controls the behavior of $\Omega$.
This choice of rotation law automatically holds the 
rough Rayleigh stability condition i.e., the specific angular 
momentum ($j$) should not become smaller outwards.
Two limiting cases are $A\to\infty$ and $A\to0$.
The former choice gives the uniform rotation and the latter 
one leads to rotations with a constant angular momentum (KEH).

By integrating equation (7) with the rotation law given in equation (9), 
we get the integral representation of the hydrostatic equation as follows,
\ba
\ln H +
\frac{1}{2} \ln \left( 1+2\Phi \right) \nonumber \\
~~ +\frac{1}{2} \ln \left( 1-v^2 \right) -\frac{1}{2} A^2 \left( \Omega - \Omega_C \right)^2 = C ,
\ea
where $C$ is an integration constant.
By imposing boundary conditions, this equation can be used to obtain 
the rest mass density. From equations (1), (2), (4) and (8) 
the active mass density is given by
\be
\rho_{{\rm{active}}} = \rho_0 H \frac{1+v^2}{1-v^2} +2P .
\ee
The gravitational potential is found by solving the modified Poisson's 
equation [eq. (3)].

\section{Method}
\subsection{Hachisu's Self-consistent Field}
In addition to the integral representation of the usual SCF methods, 
\cite{hac86a} took a unique way to impose the input parameter for 
rotation. It uses an axis ratio, $r_p/r_e$ rather than $\Omega_C$.
If the equilibrium stars have no genus, i.e., their shapes are 
spheroidal or quasi-toroidal, $r_p$ and $r_e$ are the distances to 
the boundary positions, $P$ along the polar axis and $Q$ along the 
equatorial axis, respectively. Then, the ratio lies in the range 
$0<r_p/r_e<1$. For the toroidal (donut-shaped) stars, these two
distances are the equatorial distances to the inner and the outer 
boundaries respectively, and the ratio is designated to have negative 
values. Actually, \cite{hac86a} found these toroidal-shaped 
``Dyson-Wong'' sequences \citep{dys84a,dys84b,won74} very
accurately and much more efficiently compared to the previous work 
by \cite{eri81}. Note this is not a simple task if $\Omega_C$ would 
have to be chosen as the input parameter, since there 
could be a degeneracy in the rotation speed for a given angular 
momentum \citep{hac86a}.

Our approach is almost the same as the one taken  by KEH except 
for having only one gravitational potential to solve. The major 
difference from the Newtonian approach is that $r_e$ appears explicitly
in the equations and should be determined during the numerical 
calculation. 
If one finds $r_e$ and $\Omega_C$, the distribution of the angular 
speed $\Omega$ and $v$ can be obtained.
If the integration constant $C$ is additionally known, a final
equilibrium solution can be found from equation (10) for a given $\Phi$.
These steps are repeated iteratively as described below in detail.
Note that we recover the uniform rotation if we choose 
$A^2\rightarrow \infty$ while $A=0$ corresponds to the constant 
specific angular momentum.

\subsection{Determination of $r_e, \Omega$, and $C$}
To determine $r_e, \Omega_C$ and $C$, we need to solve equations
(9) and (10) at three points. Two of them are the two boundary 
positions $P$ and $Q$. The third one is $W$ where the enthalpy 
or $\rho_0$ has its maximum value. Then we have six equations for 
the six unknowns $\Omega_P, \Omega_Q, \Omega_W, r_e, \Omega_C$, 
and $C$ as follows,
\ba
&&\frac{1}{2} \ln \left( 1+2 \Phi_P \right) + 
\frac{1}{2} \ln \left( 1-v_P^2 \right) - \frac{1}{2} A^2 
\left( \Omega_P - \Omega_C \right)^2 = C,  \, \nonumber \\
&&\frac{1}{2} \ln \left( 1+2 \Phi_Q \right) + \frac{1}{2} \ln 
\left( 1-v_Q^2 \right) - \frac{1}{2} A^2 
\left( \Omega_Q - \Omega_C \right)^2 = C,  \, \nonumber \\
&&\ln H_{\rm{max}} + \frac{1}{2} \ln \left( 1+2 \Phi_W \right) + 
\frac{1}{2} \ln \left( 1-v_W^2 \right)  \nonumber \\
&&~~~~ ~~- \frac{1}{2} A^2 \left( \Omega_W - \Omega_C \right)^2 = C,  \, \nonumber \\
&&A^2 \left(\Omega_C-\Omega_P\right)= \frac{\Omega_P r_P^2 \sin^2\theta_P}
{\left(1+2 \Phi_P\right)^2-\Omega_P^2 r^2 \sin^2\theta_P}, \, \nonumber \\
&&A^2 \left(\Omega_C-\Omega_Q\right)= \frac{\Omega_Q r_Q^2 }{\left(1+
2\Phi_Q\right)^2-\Omega_Q^2 r_Q^2 }, \, \nonumber \\
&&A^2 \left(\Omega_C-\Omega_W\right)= 
\frac{\Omega_W r_W^2}{ \left(1+2\Phi_W\right)^2-\Omega_W^2 r_W^2 } .
\ea
Since the energy density vanishes at $P$ and $Q$, we have 
imposed two boundary conditions, $H(P)=H(Q)=1$.
Note that we use $\rho_{0}^{\rm{max}}$ as a free parameter, 
so that $H_{\rm{max}}$ is known already.
We solve these equations by using the Newton-Raphson method.
For the spheroidal and the quasi-toroidal stars, the number of 
unknowns is reduced to four because two unknowns are predetermined 
by the relations, $\Omega_P = \Omega_C$ and $C=\frac{1}{2} 
\ln \left( 1+2 \Phi_P \right) $. Moreover, for the uniformly 
rotating stars the relations can be further reduced,
since angular speed is the same at all the positions 
($\Omega_P = \Omega_Q = \Omega_W = \Omega_C$). In this case,
the values of $r_e$ and $C$ are easily determined in closed form.

\subsection{Calculation of $\Phi$}
In the spherical polar coordinates, the gravitational potential 
$\Phi$ is given by
\ba
\Phi\left(r\right) &=& -G \int \frac{\rho_{\rm{active}}}{\left|r'-r\right|} d^3 r \nonumber \\
&=& - 4\pi G \int_{0}^{\infty} dr \int_{0}^{1} d\mu \nonumber \\
&&\times \sum_{n=0}^{\infty} f_{2n}(r',r) P_{2n}(\mu) P_{2n}(\mu') \rho_{\rm{active}} ,
\ea
where $\mu=\cos\theta$, $P_{2n}$ is Legendre polynomial of 
order $2n$ and $f_{2n}$ is defined as
\be
f_{2n} = \left\{ \begin{array}{ll}
r'^{2n+2}/r^{2n+1} & \textrm{if   } r' < r \\
r^{2n}/r'^{2n-1} & \textrm{if   } r' >  r
\end{array} \right. .
\ee
It is much more efficient to perform the integration expressed 
with the Legendre polynomials than taking a direct 3-volume integration.
The integration is carried out by using the Simpson's 
formula with second order accuracy.

Sometimes, we used the overrelaxation technique to improve the 
convergence \citep{var62}. Then, the gravitation potential is a 
weighted average of a newly calculated value
and the one in the previous iteration step.
That is
\be
\Phi^{n+1} = w \Phi^{n+1} + (1-w) \Phi^n ,
\ee
where $\Phi^{n}$ and $\Phi^{n+1}$ are the gravitational 
potential at $n$-th and ($n+1$)-th iteration steps, respectively.
The weight factor $w$ is determined empirically, so its value 
is problem-dependent. We used the value $w=1$ for usual cases 
(no weight) and $w=0.6$ for not easily converging problems.

\subsection{Solution procedure}
The computational grid covers from $r=0$ to $r=\frac{16}{15} r_e$ 
which is the same as Hachisu's original one but different from 
KEH who used $r_{\rm{max}}=2r_e$. For the initial model of 
non-rotating star before the iteration procedure,
we use the modified Toleman-Oppenheimer-Volkoff (TOV) equation 
adapted to our configuration. Since this equation is ordinary 
differential equation that depends only on $r$, we solve it by
using the fourth order Runge-Kutta method.
we used the iterative method for the
determination of the gravitational potential at the center
since we do not know the exact value of it.
After getting the non-rotating model, we decrease the axis ratio 
for the rotating stellar model. The following is a brief summery 
of iteration procedure: \\

\noindent 1) Initial guess of the non-rotating stellar model from modified 
TOV equation \\
2) Calculation of the gravitational potential $\Phi$ from 
density distribution \\
3) Determination of the value of equatorial radius ($r_e$), 
rotation velocity at the axis ($\Omega_C$), at the boundary and 
maximum density position 
($\Omega_{P}$, $\Omega_{Q}$, $\Omega_{W}$), and integration constant 
($C$) using Newton-Raphson method. See equation (12) \\
4) Calculation of enthalpy ($H$) from the value obtained in 3) 
and equation (9)\\
5) Conversion of enthalpy ($H$) to the rest mass density 
($\rho_0$) from equations (5) and (6)\\
6) Go back to step 2) until $\max \left| \frac{\Phi^{\rm{new}}-
\Phi^{\rm{old}}}{\Phi^{\rm{new}}} \right| < \delta $, $\max \left| 
\frac{H^{\rm{new}}-H^{\rm{old}}}{H^{\rm{new}}} \right| < \delta $, 
and $ \left| \frac{C^{\rm{new}}-C^{\rm{old}}}{C^{\rm{new}}} \right| 
< \delta $, where $\delta$ is desired accuracy
(iteration criteria). We have chosen $\delta$ to be $10^{-12}$ \\

The calculation of the ring-like toroidal solution starts at axis 
ratio=-0.8. The initial guess of the ring solution comes from 
the Newtonian solution to prevent from the failure in getting 
solutions using Newton-Raphson method while solving equation (12),
because its solution procedure sensitively depends on its initial guess.
After calculating ring solution with axis ratio=-0.8, we increase 
the axis ratio to get a solution with different axis ratio.

As for the spheroidal solutions, we start with $r_p/r_e=1$.
All the procedures are carried out until mass shedding occurs 
when the centrifugal force is so large that the gravitational force 
cannot overcome it. In that case, no stable solution of the 
rotating star exists.

\subsection{Units}
In this paper, we use the units of $c=G=M_{\odot}=1$.
With this choice, the length, mass and time units can be 
automatically determined. Consequently, the units of 
$\kappa$ and $\rho_0$ can also be determined. 
Note that the unit of $\kappa$ depends on the
choice of the polytropic index $N$. In Table 1, we summarize the 
units of these quantities. Our figures are usually expressed in these
units.

\begin{table}
\caption{Units of some physical quantities when $c=G=M_{\odot}=1$}
\begin{tabular}{cc}
\hline\hline
 Parameter & Unit Physical Scale in cgs \\
\hline
Length & $1.47\rm{km}$ \\
Time & $4.92 \times 10^{-3}\rm{ms}$ \\
Mass Density & $6.26 \times 10^{17} \rm{g}/\rm{cm}^3$\\
Frequency (1/Time) & $2.03\times 10^5 \rm{Hz}$\\
\hline\hline
\end{tabular}
\end{table}

\section{Comparison with Other Methods}
In this section, we compare our result 
(pseudo-Newtonian Relativistic hydrodynamics approach, pNRH hereafter)
with three other methods: purely Newtonian, Newtonian relativistic 
hydrodynamics (NRH hereafter) in that special relativity is taken 
into account with Newtonian gravity, and general relativistic approaches
(GR hereafter). To obtain the general relativistic solutions, 
we use the Whisky\_RNSID code in Whisky project 
(http://www.whiskycode.org). The code can calculate the equilibrium 
solutions of rotating stars  for various equations of states. 
We have not implemented the 
toroidal configuration in the Whisky code yet.
Hence, we compare only spheroidal (uniform rotation) and 
quasi-toroidal (differential rotation) stellar solutions.

For the future reference, we introduce the following quantity that 
measures the importance of the relativistic effects,
\be
R=H^{\rm{max}}-1=(N+1)\kappa\left(\rho_{0}^{\rm{max}}\right)^{1/N} ,
\ee
which could be comparable to or larger than unity if the relativistic 
effects are important and $R \ll 1$ for the non-relativistic case.
Another direct measure of the importance of the general relativistic
effects is the ration between the gravitational potential measured in units
of $c^2$, i.e., 
\be
\psi = {GM \over c^2 r_{eff}},
\label{psi}
\ee
where $r_{eff}$ is the radius of a sphere whose volume is identical
to the real volume of the star. On the other hand, 
the special relativistic importance due to rapid rotation can be measured
by the dimensionless parameter
\be
\chi = {Jc \over GM^2},
\label{chi}
\ee
where $J$ is the angular momentum of the star. Unlike $R$, $\psi$ and $\chi$
should always remain to be less than unity. The relativistic effects become
significant if these paramegers are not very small compared to unity. 
The angular momentum can be computed by
\ba
J &=& \int d3 x T^{t}_{\phi} \sqrt{-g} \nonumber \\
&=& \int d3 x \rho_0 H \frac{v2}{1-v2} \frac{1}{\Omega} \sqrt{-g},
\ea
where $g$ is the determinant of the metric.

\subsection{Uniform Rotation}

The density profiles obtained by these four different methods for
uniformly rotating star with axis ratio of 0.7 along the equatorial 
radius are shown in Figure 1. For this model we used $N=1$, $\kappa=100$, 
and $\rho_{0}^{\rm{max}}=0.001$, so that $R=0.2$.

\begin{figure}
\includegraphics[width=84mm]{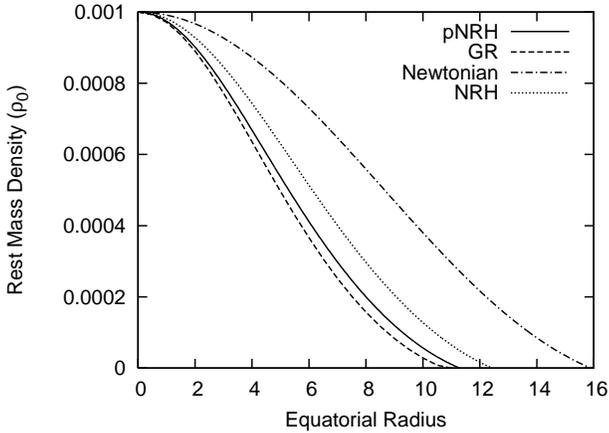}
\caption{Rest mass density of the uniformly rotating stellar models
with the axis ratio $r_p/r_e=0.7$ at the equatorial plane obtained by 
four different approaches. Our proposed method of pNRH (solid) provides
solution that agrees well with the 
general relativistic solution (dashed). The equatorial radius of
the Newtonian result (dot-dashed) is larger by about $50\%$ 
compared to the general relativistic result as well as our one.
For comparison, we also have plotted the result from NRH (dotted).
Evidently, the introduction of the active mass makes further 
improvement of NRH. Note that $\psi=0.53$ and $\chi=0.17$ for the
model shown here, when these quantities are computed based on
our pseudo-Newtonian (pNRH) approach.
}
\end{figure}

It is evident that our approach of pNRH gives the density profile
very similar to that of the GR solution. Our pNRH result for the 
equatorial radius agrees with that of general relativity within 
5\%. On the other hand, the Newtonian solution is significantly
different from the general relativistic one.
For example, the the equatorial radius of 
Newtonian solution is about 50 \% larger than of the GR method. 

NRH that does not take into account the active mass gives the solution 
closer to the GR one compared to the Newtonian approach, but we can
see further improvement with pNRH. 
Note the value of $R=0.2$ leads to rather small relativistic correction.
The relative accuracy of our pseudo-Newtonian 
approach over the NRH would be 
even more appreciable for models with larger $R$.
Also note that $\psi=0.17$ and $\chi=0.53$ for the
model shown here, when these quantities are computed based on
our pseudo-Newtonian (pNRH) approach.

\subsection{Differential Rotation}
\begin{figure}
\includegraphics[width=84mm]{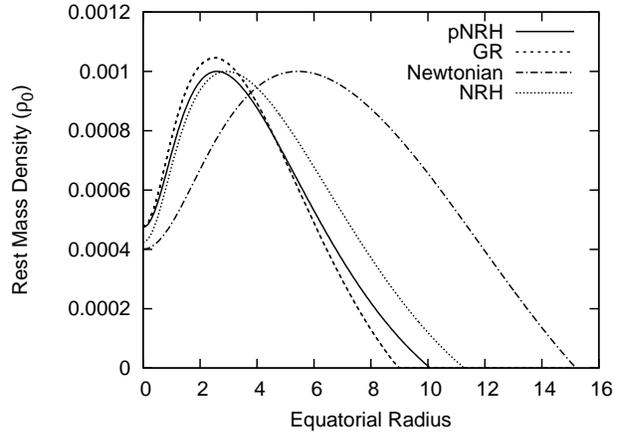}
\caption{Same as Figure 1 for the differentially rotating star 
with the axis ratio $r_p/r_e=0.35$. The shape of such model
is quasi-toroidal so that the maximum density does not
occur at the center (see \S 4.2 for further details).
The similar trend of better 
agreement of pNRH with GR than between Newtonian or NRH with GR
is evident from this figure. For this model, we found that 
$\psi = 0.29$ and $\chi =0.57$.
}
\end{figure}

Figure 2 shows the quasi-toroidal density profiles of differentially 
rotating model with $r_p/r_e = 0.35$.
We took the same parameter set with Figure 1 except for 
$\rho_{0}^{\rm{max}}$ and $A$.
Since the general relativistic code, Whisky\_RNSID uses the 
central rest mass density $\rho_{0}^{\rm{center}}$ as an input 
parameter rather than the $\rho_{0}^{\rm{max}}$,
we used the value of $\rho_0^{\rm{center}}$ of the GR solution with 
given $\rho_{0}^{\rm{max}}$ as an input parameter for
the pNRH. The code also uses a rescaled $\hat A = A/r_e$, so we had 
to find a proper $\hat A$ which could make $A^2=10$ used in the figure.
The parameters that measure the importance of the relativistic effects
are found to be $\psi =0.29$ and $\chi=0.57$. 

Similar trend with the uniformly rotating model of
Figure 1 is seen in differentially rotating model of Figure 2: 
progressive improvement from Newtonian approach toward the
true solution of GR through NRH and pNRH. 
Our result of pNRH (solid) for the density distribution along the major 
axis shows good agreements with the GR solution (dashed line). 
The location of maximum density of pNRH differs by $5\%$ 
compared to that of GR. 
Newtonian solution (dot-dashed) has the 
equatorial radius which is about two times larger than that of
GR one. The location of the maximum density
is even more different. 
The difference in these quantities between pNRH and GR is much smaller 
than the difference between GR and Newtonian or between GR and NRH.
Note that the maxumum densities are similar among
pNRH, NRH and Newtonian, but GR value is slighter larger.

Not shown in the figures, we have calculated the equilibrium 
solutions for the Newtonian hydrostatic equation coupled with 
the modified Poisson's equation that takes into account the active mass.
The solutions are not much different from the purely Newtonian 
case since the active mass differs from the rest mass only slightly.
We conclude that the relativistic treatment of the hydrodynamics 
is crucial for better agreement. Also, it could be improved 
significantly by taking into account the active mass density for 
relativistic cases.

\section{Characteristics of the Pseudo-Newtonian Relativistic Solutions}
Newly born neutron stars rotate differentially and are near the 
critical rotation, but turn into uniformly rotating stars in
relatively short time scale because of the shear viscosity and the 
magnetic tensions.
This means that old neutron stars tend to rotate uniformly.
We now present the detailed characteristics of the pNRH solutions
for rotating compact stars. We discuss our results separately for
uniformly and differentially rotating cases.

\subsection{Uniform Rotation}
\subsubsection{Spheroidal and Toroidal Solutions}
Examples of our equilibrium solutions for the uniformly rotating stars
are shown in Figure 3.
The top and bottom panels show a spheroidal and a 
toroidal solutions, respectively.
We took $\rho_{0}^{\rm{max}} =0.001$ for the spheroidal star in the top 
panel. This choice of $\rho_{0}^{\rm{max}}$ corresponds to physical 
density of a few times larger than the nucleon density 
of $2.7 \times 10^{14} \rm{g}/\rm{cm}^3$.
Even though the polytropic equation of state is not realistic, 
these values for $\kappa, N$, and $\rho_{0}^{\rm{max}}$ are
widely used because this set produces a typical size and a rest mass 
density of neutron stars. Our results show that the physical size 
is about 16.6 km which is $\sim 5\%$ larger than the general 
relativistic result.
With the axis ratio $(r_p/r_e)=0.7$, the angular frequency is 
calculated to be 597 Hz. For the toroidal star, we take a ten times 
smaller density, $\rho_{0}^{\rm{max}}=0.0001$ which produces a star 
with a much larger size. We note here that the axis ratios for toroidal
models are denoted by negative numbers following the convention by
Hachisu (1986ab). The axis ratio is selected to be -0.35.
The size and the rotation frequency obtained are 49.6 km and 199 Hz, 
respectively. The critical rotation occurs at 
$r_p/r_e= 0.525$ $(f_{rot} = 684 {\rm{Hz}})$ for the spheroidal star 
and at $r_p/r_e= -0.204$ $(f_{rot} = 228{\rm{Hz}})$.
\begin{figure}
\includegraphics[width=84mm]{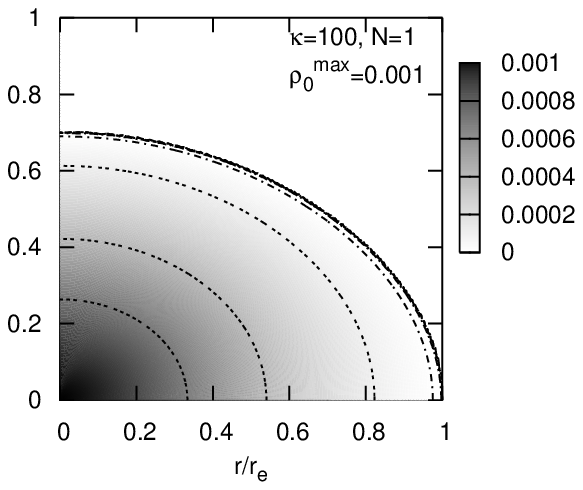}\\
\includegraphics[width=84mm]{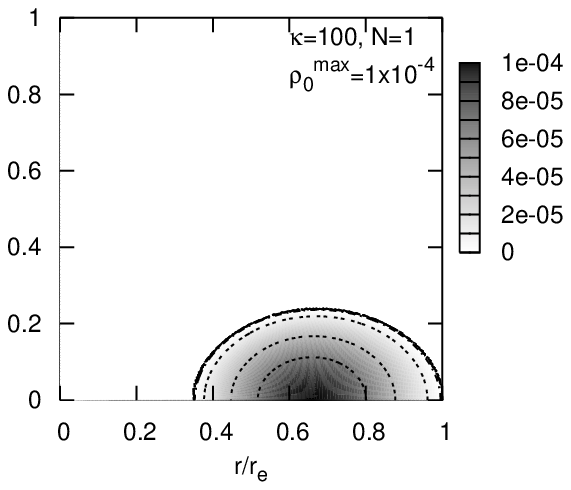}
\caption{The distribution of $\rho_0$ of a spheroidal star (top) and a 
toroidal star (bottom) for $\kappa=100$, and  $N=1$.
For the spheroidal star, $\rho_{0}^{\rm{max}}=0.001$ and 
$r_p/r_e=0.7$ are used as input parameters, which are the same
for the model shown in Fig. 1. 
The physical size $(r_e)$ and the rotation frequency ($f_{rot}$) are
16.6 km and 597 Hz, respectively. For the toroidal star, 
$\rho_{0}^{\rm{max}}=0.0001$ and
$r_p/r_e=-0.35$. The physical size is 49.6 km and the rotation
frequency is 199 Hz.}
\end{figure}

As expected, our solutions show typical features of polytropic 
rotating stars. If $\kappa$ increases, the size of star increases 
and the rotation speed decreases. The density falls off more rapidly, 
the size becomes bigger, and the rotation speed becomes smaller as  
$N$ increases.

\subsubsection{Relation between axis ratio and angular velocity}
In Figures 4, we show equilibrium solution sequences of angular speed 
as a function of ellipticity which is defined as 
$\varepsilon = 1 - r_p/r_e$, for different choices of 
$\rho_{0}^{\rm{max}}$'s, represented by different line
types, and three differnt values of polytropic indices ($N$), 
separated by different panels.
The values of $\kappa$ are differnt for different choices
of $N$: $N=1$ and $\kappa=100$ for the top, $N=1.5$ and $\kappa=10$ 
for the middle, and $N=3$ and $\kappa=0.1$ for the bottom panel.
Each panel has three sequences which are three different 
values of $\rho_{0}^{\rm{max}}$: $10^{-2}$, $10^{-3}$, and $10^{-4}$.
Note that the values of $R$ for the 
$\rho_{0}^{\rm{max}}=10^{-2}$, $10^{-3}$ and $10^{-4}$ are 
2, 0.2, and 0.02 for $N=1$, and $R=1.16$, 0.25 and 0.054 for 
$N=1.5$ and $R=0.086$, 0.04 and 0.019 for $N=3$ respectively.

As shown in the figures, $\Omega_C$ increases with 
$\rho_{0}^{\rm{max}}$ and the sequences terminates at the 
critical rotation speed beyond which the mass shedding occurs.
For the Newtonian case, due to the simple scaling property of  
$\Omega_C \propto \sqrt{\rho_{0}^{\rm{max}}}$, the 
critical rotation always occurs at the
same $r_p/r_e$ \citep{hac86a}.
However, our model shows somewhat different results. 
For the cases of $N=1$ and $N=1.5$, high density models are relativistic
but low density models are not. The scaling relation of $\Omega_C\propto 
\sqrt{\rho_0^{{\rm max}}}$ does not hold if the models are relativistic.
However, we found that the models with $N=3$ are all non-relativistic, and
three different models with different $\Omega_C$ follows the Newtonian
scaling relationship very well. The maximum value of ellipticity (or minimum
values of axis ratio) are also nearly the same for different $\rho_0^{max}$.

Also shown in the inlets of each panel are the $\Omega_C/\Omega_{C,max}$
versus $\varepsilon/\varepsilon_{max}$ where the maximum values of $\Omega_C$ and
$\varepsilon$ correspond to the critically rotating models. 
In such normalizations,the behavior of $\Omega_C$ with $\varepsilon$ is almost
identical for models with different density if the polytropic index is fixed.
For different $N$, however, the behaviour of $\Omega_C$ becomes somewhat 
different. Thus we conclude that the scaling relationship of $\Omega_C$ with
$\varepsilon$ depends on the degree of relativity while the functional
behaviour is mostly determined by the equation of state.



\begin{figure}
\includegraphics[width=80mm]{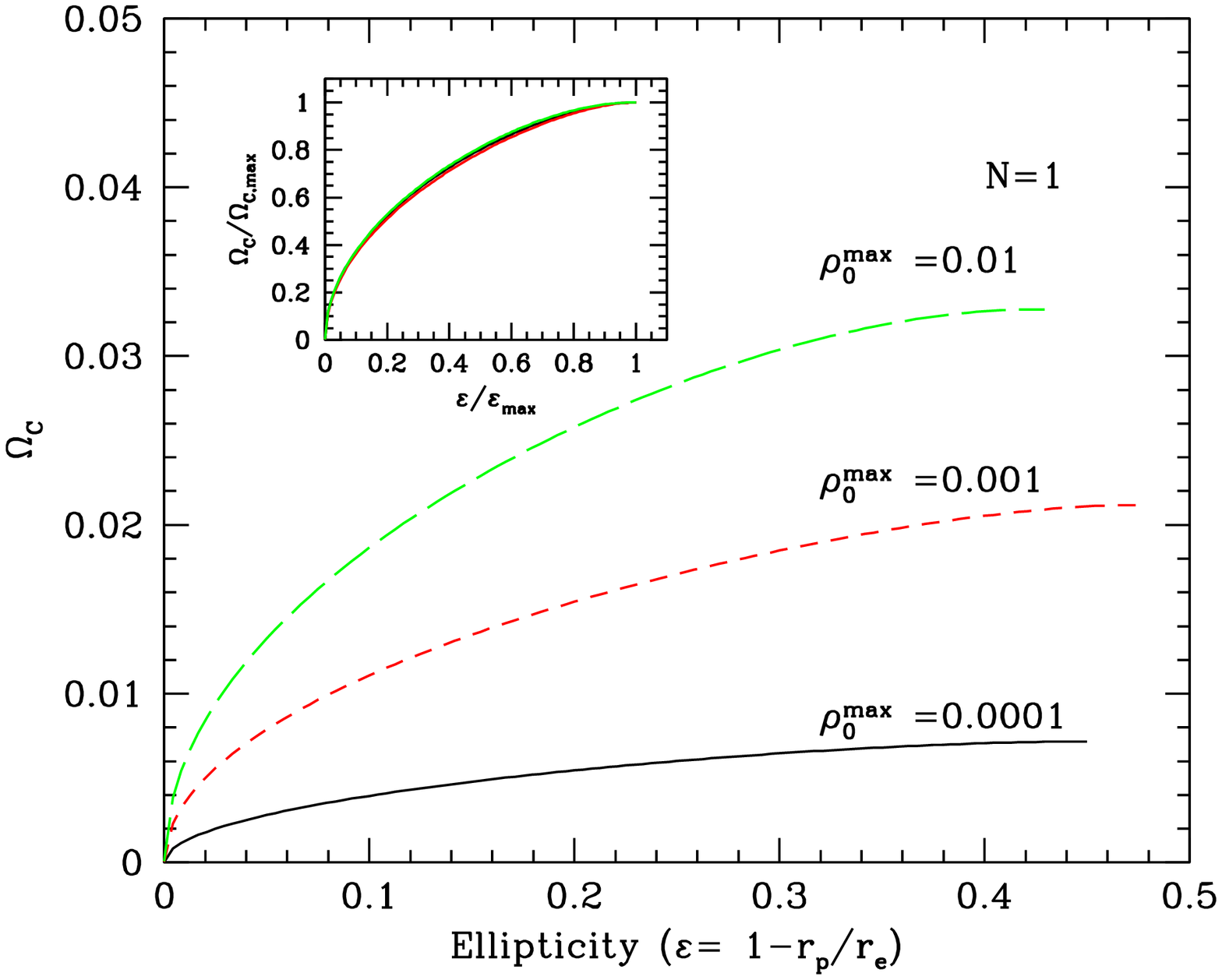}\vspace{-2cm}\\
\includegraphics[width=80mm]{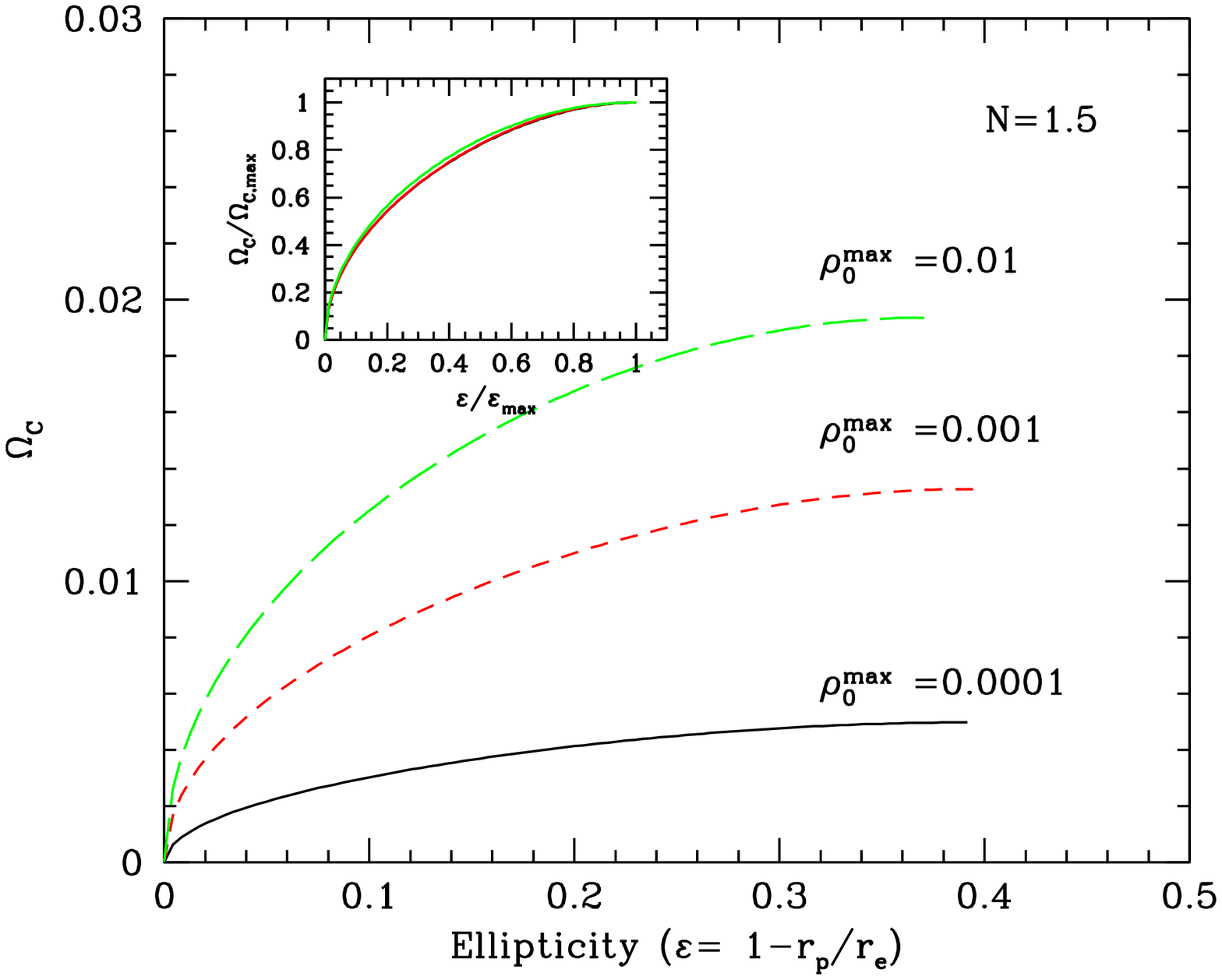}\vspace{-2cm}\\
\includegraphics[width=80mm]{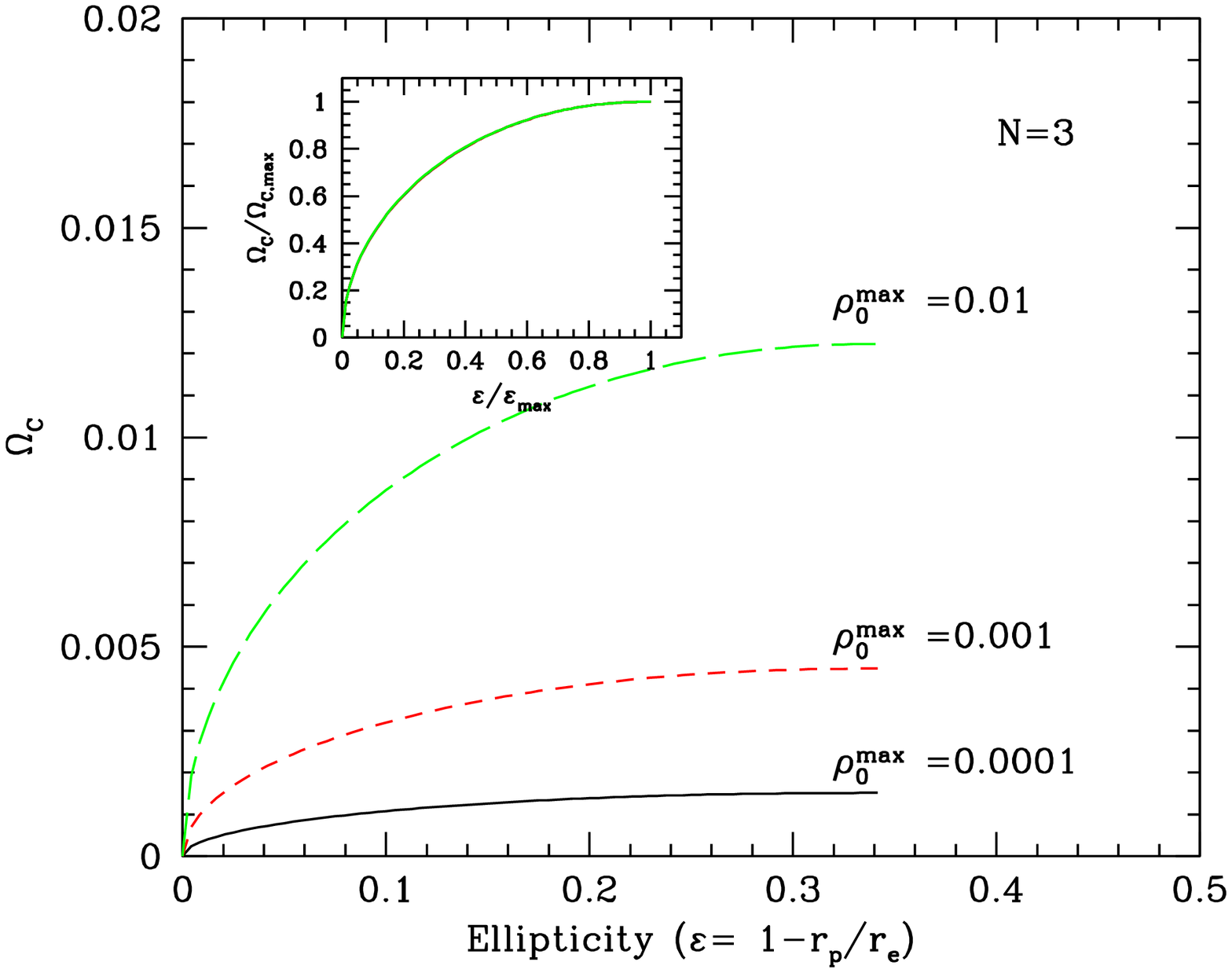}\vspace{-1cm}
\caption{
The rotation angular speed at the rotation axis ($\Omega_C$) 
versus the ellipticity for 
$N=1 ~(\Gamma=2)$, $N=1.5 ~(\Gamma=5/3)$, and $N=3 ~(\Gamma=4/3)$ 
from top to bottom panels.
The three lines in each panel represent the results with 
$\rho_0^{\rm{max}}=0.01$, $0.001$ and $0.0001$ from the top.
The corresponding values of $R$ defined in equation 
(16) from the top are:
$R=2$, 0.2, and 0.02 for $N=1$, and 1.16, 0.25, and 
0.054 in case of $N=1.5$, and 0.086, 0.040 and 0.0186, respectively.
We also show the normalized angular speed versus normalized ellipticity
in the inlet of each panel. These figures show that the functional
behaviour of the angular speed at the rotational axis mostly depends
on the equation of state.
}
\end{figure}

\subsubsection{Constant Rest Mass Sequences}
Most of the neutron stars have their masses near the 
Chandrasekhar limit ($1.4M_{\odot}$).
Even if they undergo dynamical changes, the baryon number of the 
star is conserved. Hence, the sequences in fixed rest mass can give 
an insight into the neutron stars' dynamical properties.
The rest mass of the star is obtained by
\ba
M_0&=&\int \rho_0 u^t \sqrt{-g} d^3 x \nonumber \\
&=&2\pi \int dr \int d\theta r^2 \sin\theta \frac{\rho_0}{\sqrt{1-v^2}} 
\left(1+2\Phi\right)^{-\frac{3}{2}}.
\ea
In order to obtain the constant mass models, we
adjusted the maximum density while keeping the equation of 
states unchanged.

\begin{figure}
\includegraphics[width=84mm]{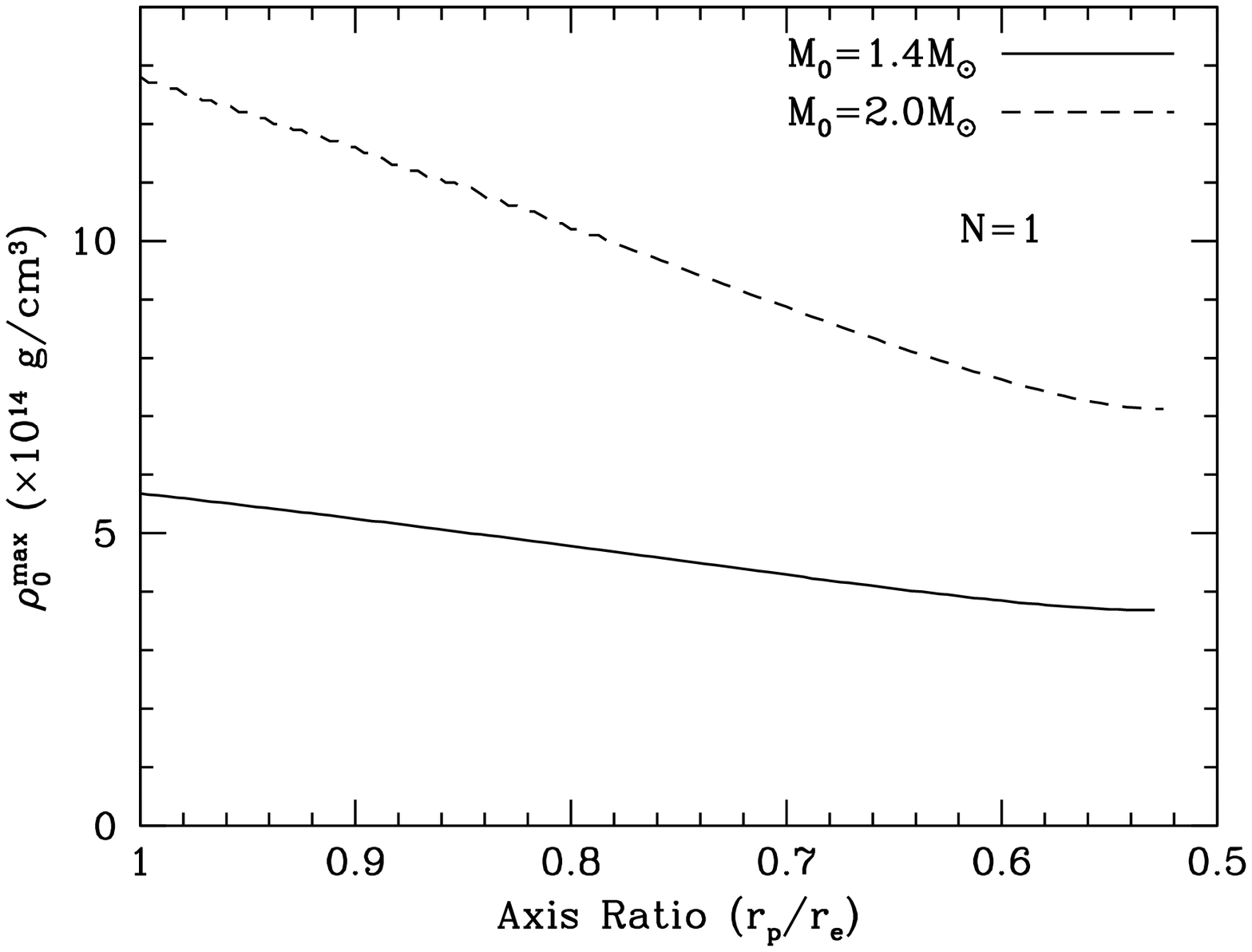}\vspace{-2cm}\\
\includegraphics[width=84mm]{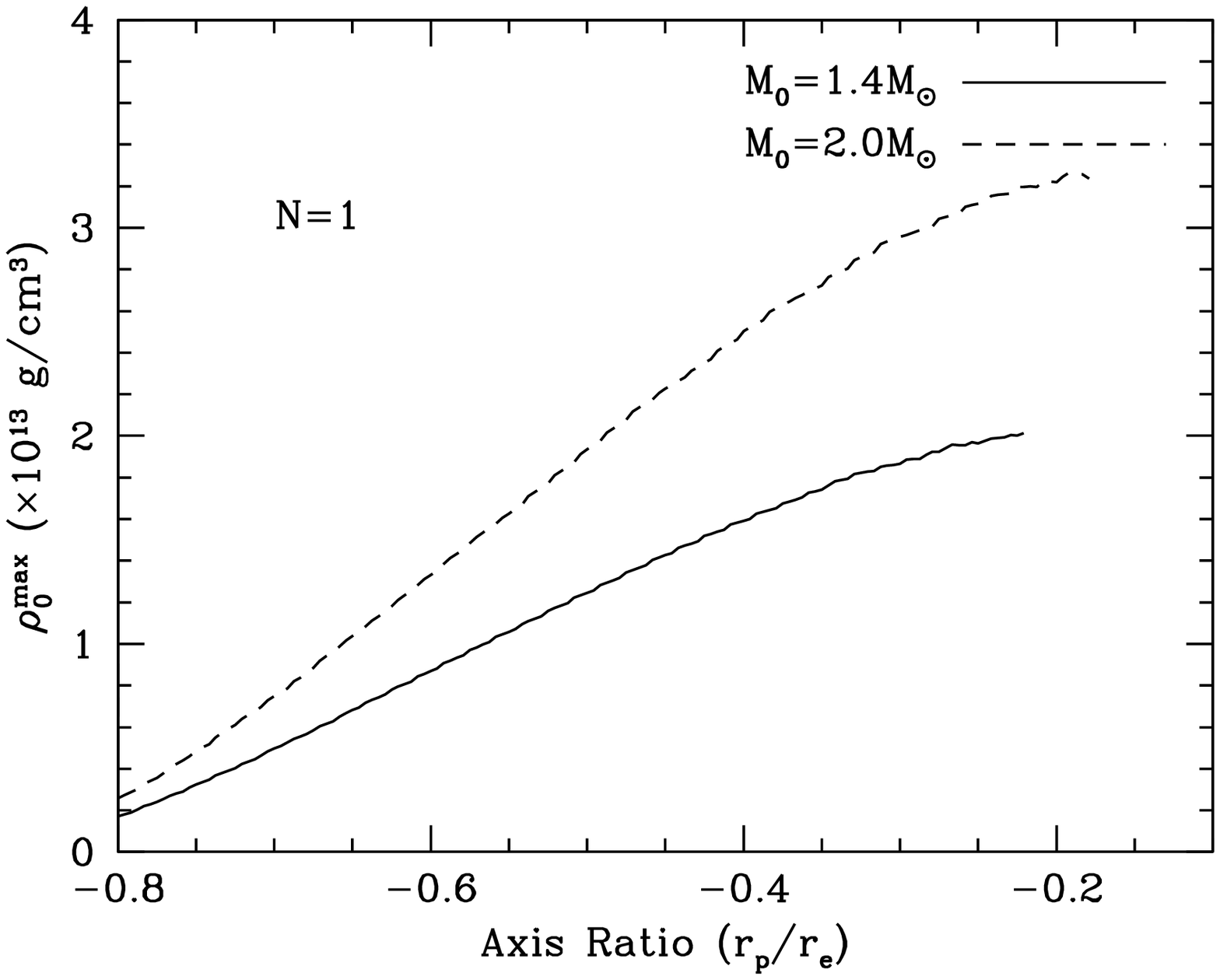}\vspace{-1cm}
\caption{Axis ratio ($r_p/r_e$) - maximum rest mass density ($\rho_{0}^{\rm{max}}$) relation in constant masses ($1.4M_{\odot}$ and $2M_{\odot}$) for spheroid (top panel) and toroid (bottom panel).
}
\end{figure}
Figure 5 shows the relation between the axis ratio ($r_p/r_e$) and 
maximum rest mass density ($\rho_{0}^{\rm{max}}$).
Since more elongated figures rotates faster and their sizes are larger, 
they do not need the large value of $\rho_{0}^{\rm{max}}$.
Accordingly, $\rho_{0}^{\rm{max}}$ decreases as axis ratio decreases
for spheroid and toroid. For the toroidal case, the density is much 
smaller than the spheroidal one and its size is very 
sensitively dependent on the value of $\rho_{0}^{\rm{max}}$.

\begin{figure}
\includegraphics[width=84mm]{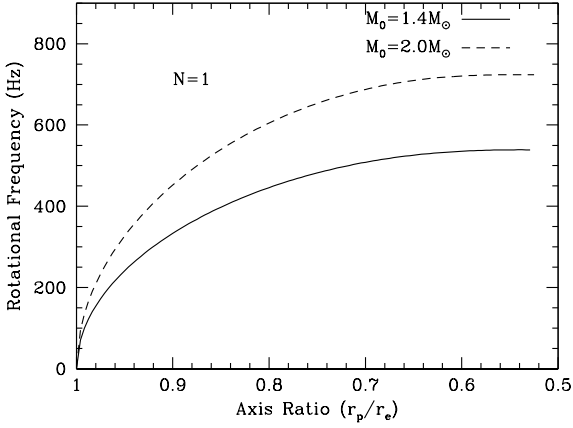}\vspace{-2cm}\\
\includegraphics[width=84mm]{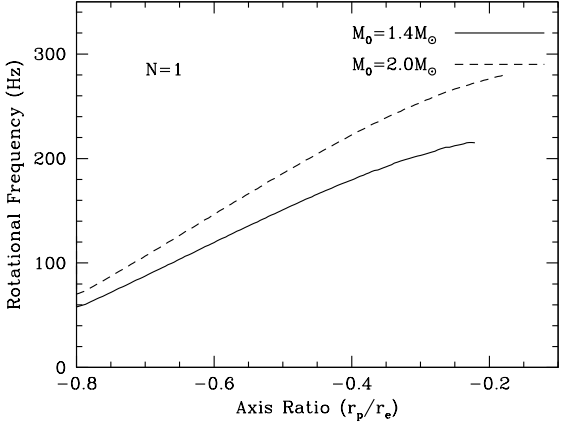}\vspace{-1cm}
\caption{Axis ratio ($r_p/r_e$) - rotational frequency 
relation in constant masses ($1.4M_{\odot}$ and $2M_{\odot}$) for 
spheroid (top panel) and toroid (bottom panel).
}
\end{figure}

The relation between the rotation frequency at the axis 
versus ratio ($r_p/r_e$) relations is shown in Figure 6 for 
spheroidal (top panel) and toroidal cases (bottom
panel).
Of course more elongated shape and massive star rotates faster  
for spheroidal shape. Rotation frequency of toroid is much smaller 
than the one for spheroid since the specific angular momentum is
larger for a given rotational frequency.
For the $1.4M_{\odot}$ spheroidal star, rotational frequency is 
539 Hz near the critical rotation and 724 Hz for the 
$2M_{\odot}$ one. For the toroidal stars, the critical 
frequencies become 215 and 280 Hz for 1.4 and 2.0 $M_\odot$, 
respectively. 

\begin{figure}
\includegraphics[width=84mm]{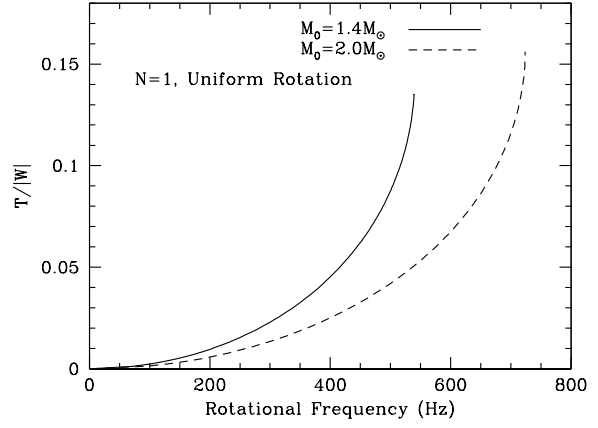}\vspace{-2cm}\\
\includegraphics[width=84mm]{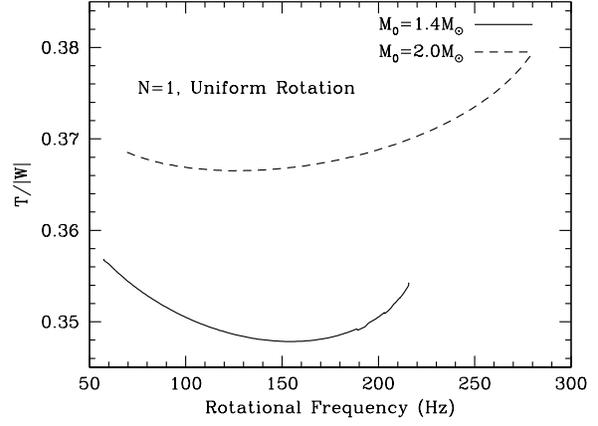}\vspace{-1cm}
\caption{Rotational frequency $\Omega$)  - $T/|W|$ relation for the
models with fixed total masses ($1.4M_{\odot}$ and $2M_{\odot}$) 
for spheroid (top panel) and toroid 
(bottom panel).
}
\end{figure}

Figure 7 shows the relation between $\Omega_C$ and the ratio of the rotational kinetic 
energy to gravitational potential energy ($\beta\equiv T/|W|$).
It is well known that a bar mode instability occurs if $\beta$
exceeds certain critical value $\beta_d$. For 
Newtonian case, the $\beta_d$ for the onset of
the bar mode instability is 0.2738 \citep{cha69,st83}.
For the case of fully relativistic 
stars, the $\beta_d$ lies between 0.24 and 0.25 \citep{sbs01} 
which is slightly smaller than Newtonian one.
The rotational kinetic energy ($T$) and gravitational 
potential energy ($W$) are defined as

\ba
T &=& \frac{1}{2} \int d^3 x \Omega T^{t}_{\phi}\sqrt{-g} \nonumber \\
&=& \frac{1}{2} \int d^3 x \rho_0 H \frac{v2}{1-v2} \sqrt{-g} ,
\ea

\be
W=T+M_p-M_{\rm{active}},
\ee
where the proper mass $M_p$ and active mass $M_{\rm{active}}$ can be obtained by
\ba
&&M_p = \int d^3 x \frac{\rho_0\left(1+\epsilon\right)}{\sqrt{1-v^2}} \sqrt{\gamma} , \\
&&M_{\rm{active}} = \int d^3 x \rho_{\rm{active}} \sqrt{-g} .
\ea
Like in Figure 6, upper panel and low panel of Figure 7 are for
spherpidal and toroidal stars, respectively. 
We find that $T/|W|$ is much smaller than the critical values 
of bar mode instability for uniformly rotating spheroids with 
masses 1.4  and 2.0 $M_\odot$. 
On the other hand, all the sequences of toroidal models have
$T/|W|$ greater than the instability criteria. Apparently, the 
toriodal configurations are all dynamically unstable. 

\subsection{Differentially Rotating Models}
\subsubsection{Solution of the Quasi-toroidal Objects}
For differentially rotating stars, the 
angular frequency at the boundary becomes much smaller than 
the uniformly rotating stars in general since the inner parts rotate
much faster than outer parts. Therefore, solutions could exist 
well below the critical rotational frequcncy for mass shedding
of the uniformly rotating stars and the resulting
shape of the star is quasi-toroidal.
In figure 8, we show an example of quasi-toroidal distribution 
of $\rho_0$ for the same parameters with the uniformly rotating 
spheroidal star in Figure 3 (i.e., $\kappa=100$, $N=1$, 
and $\rho_0^{\rm max}$=0.001), except for $A^2$ which was taken to
be 10. Also, the axis ratio was chosen to be $r_p/r_e=0.35$.
The equatorial radius for this model is 17.4 km when the mass of the
star is assumed to be 1.4 $M_\odot$.
The resulting density distribution is a quasi-toroidal one in the sense
that density does not vanish at the center. The maximum
density, however occurs at finite radius so that the
location of the maximum density is a ring, as already seen from 
Figure 2. The rotational frequency at the center is 5510 Hz, but it
reduces to about 220 Hz at the surface.  

The dependence of the rotation speed distribution on the parameter 
$A^2$ is shown in Figure 9. For $A^2=10$,  the rotation speed at 
the equatorial boundary reduces to 10\% of the one at the center.
As expected, the rotation speed falls off more rapidly as $A$ 
decreases and becomes uniform as $A$ increases.

\begin{figure}
\includegraphics[width=84mm]{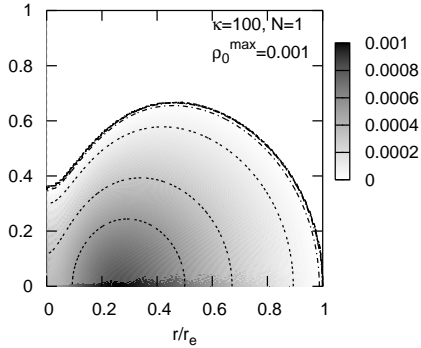}
\\
\caption{The distribution of $\rho_0$ 
for the quasi-toroidal example. 
The model parameters are the same as that of the spherdoidal 
example shown in Figure 3 (i.e., $\kappa=100$, $N=1$, 
and $\rho_{0}^{\rm{max}}=0.001$. The axis ratio of $r_p/r_e = 0.35$, 
and $A^2=10$ are chosen for the rotational parameters.  
The resulting equatorial radius is 17.4 km, when the baryon mass
is assumed to be 1.4 $M_\odot$ (see Figure 13). 
The rotation frequency at the center is 
$f_C=5510\rm{Hz}$ but it reduces down to 
4 \% of $f_C$ at the equatorial boundary.}
\end{figure}

\begin{figure}
\includegraphics[width=84mm]{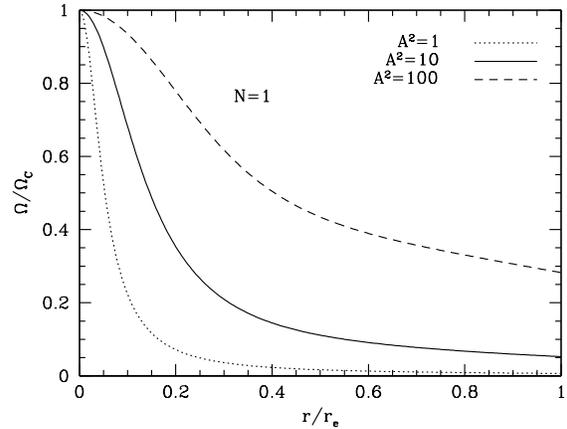}\vspace{-1cm}
\caption{The angular frequency distributions of the models with the same
parameters as that one shown in Figure 8, but three different values
of $A^2$=1, 10 and 100.
}
\end{figure}

\subsubsection{Constant Rest Mass Sequences}
In this section, we describe the characteristics of the models
with constant baryonic masses of 1.4 and 2 $M_\odot$ 
for differentially rotating objects
with $A^2$=10 and 100. Unlike the uniform rotation, critical rotation for
mass shedding does not occur in thses values of $A$.
\begin{figure}
\includegraphics[width=84mm]{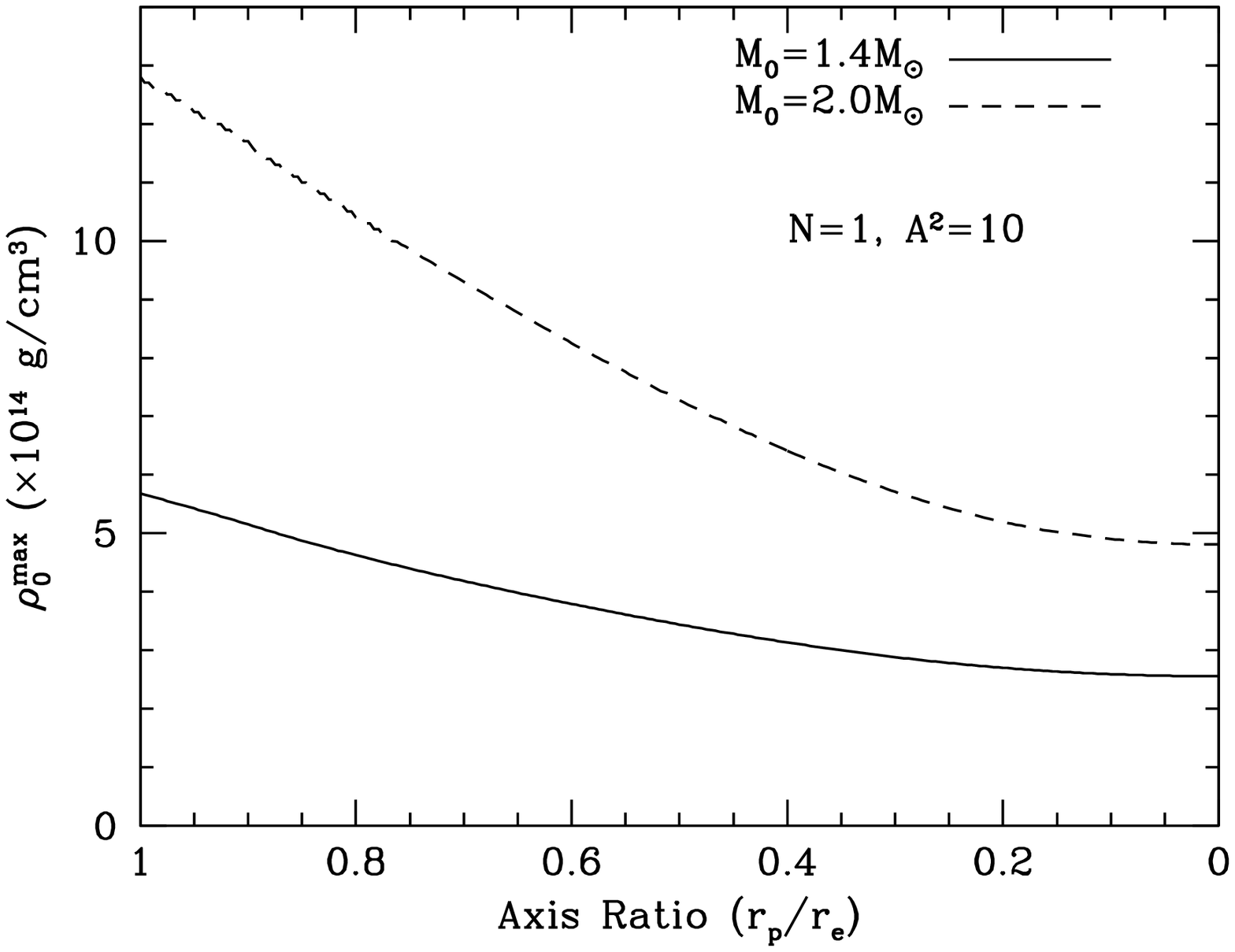}\vspace{-2cm}\\
\includegraphics[width=84mm]{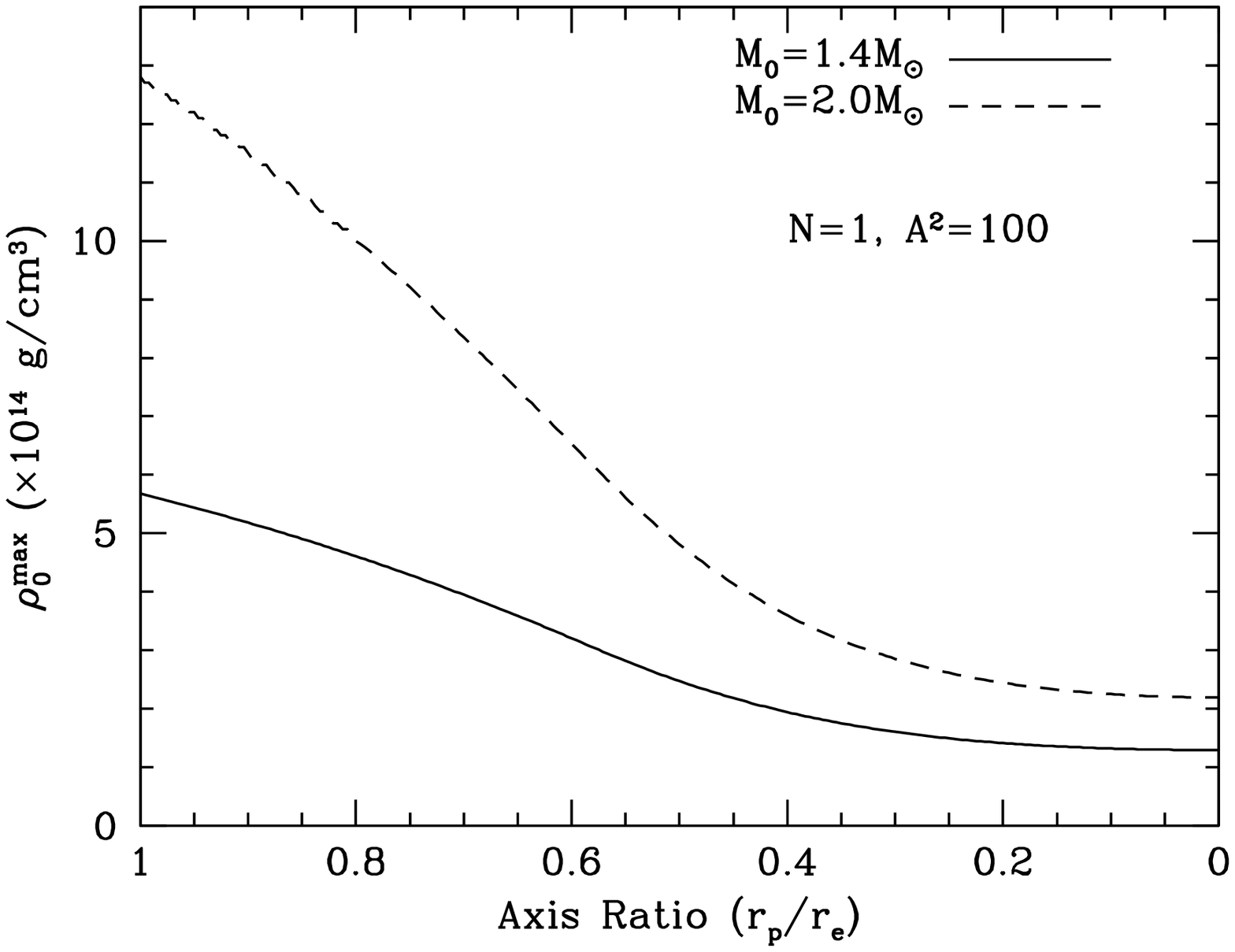}\vspace{-1cm}
\caption{ Maximum rest mass density 
($\rho_{0}^{\rm{max}}$)
versus axis ratio ($r_p/r_e$) relation for models with two
constant masses ($1.4M_{\odot}$ and $2M_{\odot}$) and
two rotational parameters of $A^2=10$ (top panel) 
and $A^2=100$ (bottom panel).
}
\end{figure}
Figure 10 shows that the relation between axis ratio 
($r_p/r_e$) and maximum rest mass density ($\rho_{0}^{\rm{max}}$).
Almost similar trend to uniformly rotating models is seen but 
$\rho_{0}^{\rm{max}}$ falls off more rapidly toward the smaller 
axis ratio. Also the maximum densities are generally larger than
the toroidal figures of uniformly rotating models, but comparable
to that of spheroidal models.

Figure 11 presents relationship between the axis 
ratio ($r_p/r_e$) and the rotational angular speed at 
the axis of rotation for the 
differentially rotating models shown in Figure 10. 
Note that the 
rotational frequency in units of Hz can be obtained by multiplying
the rotational angular speed shown in this figure by 
3.23 $\times 10^4$. Therefore, the rotational frequcncy at the axis 
of rotation could reach 
6130 Hz and 4580 Hz for 1.4 and 2 $M_\odot$ models, respectively
for the case of  $A^2=10$. For the case of $A^2=100$, the rotational
speed at the axis of rotation becomes 1520 and 1170 Hz for
1.4 and 2 $M_\odot$ models, respectively. 


\begin{figure}
\includegraphics[width=84mm]{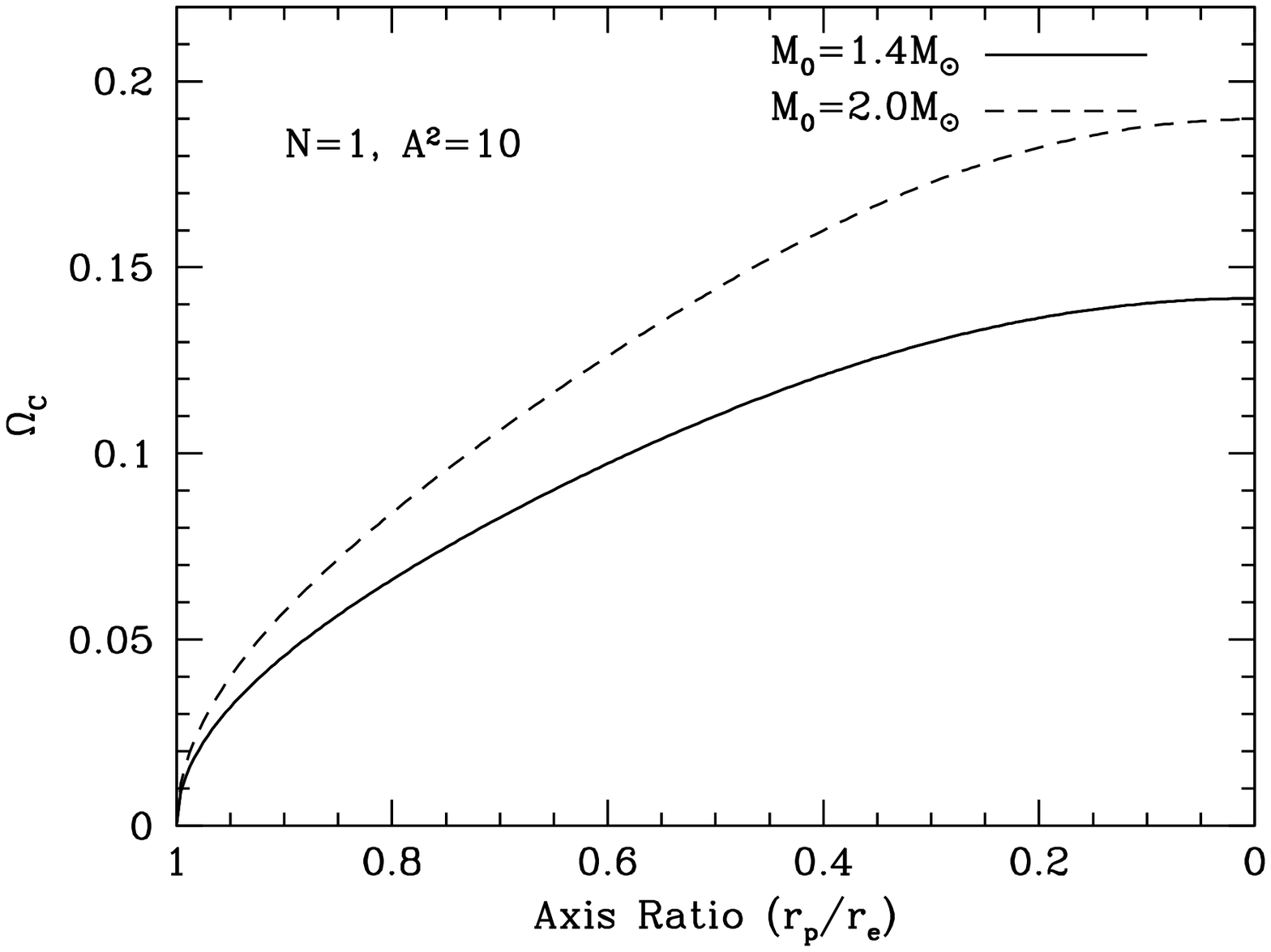}\vspace{-2cm}\\
\includegraphics[width=84mm]{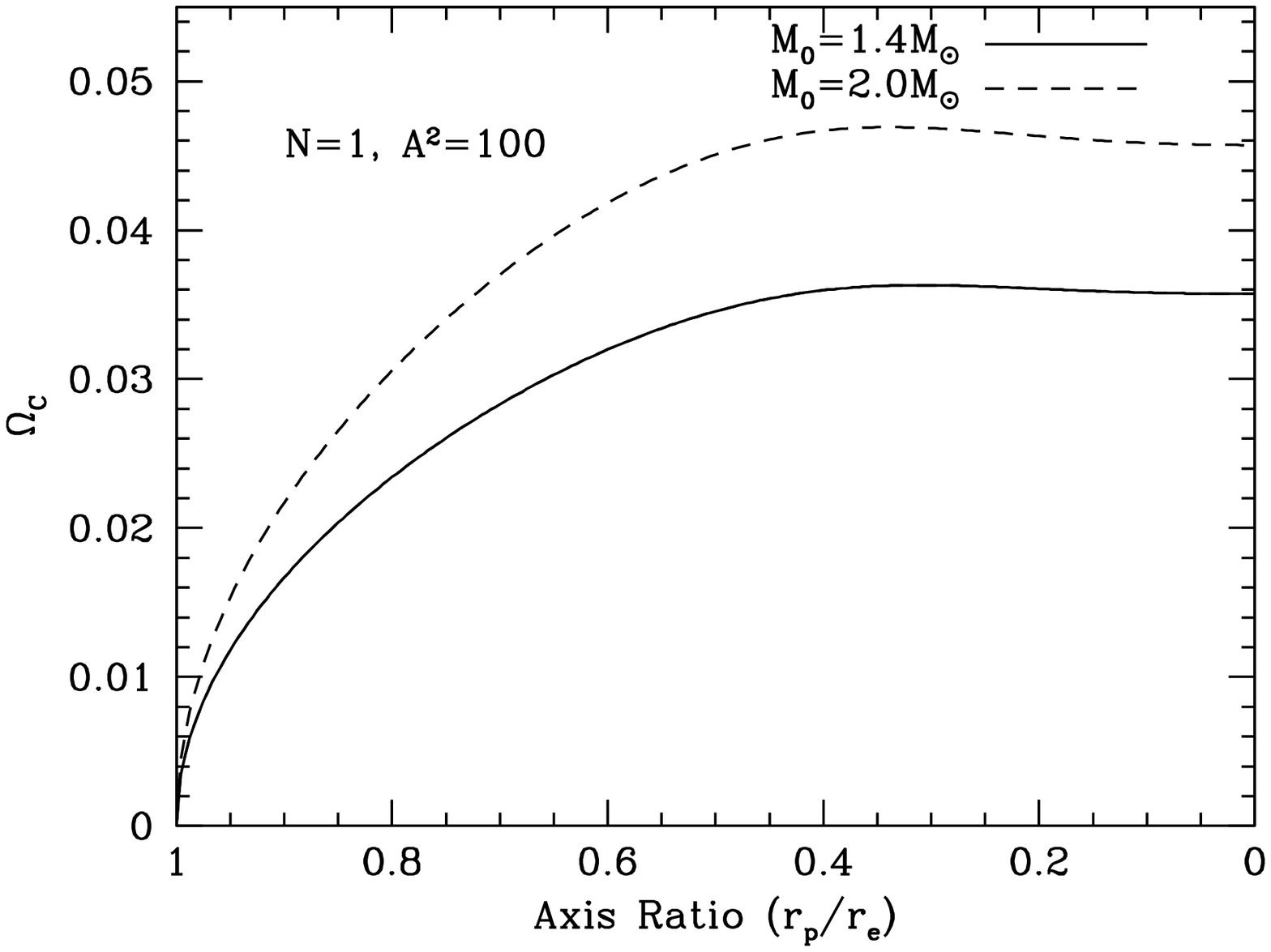}\vspace{-1cm}
\caption{Rotation angular speed along
rotation axis ($\Omega_C$) versus axis ratio ($r_p/r_e$) 
for the models with constant total masses 
($1.4M_{\odot}$ and $2M_{\odot}$) 
and differential rotation parameter $A^2=10$ 
(top panel) and $A^2=100$ (bottom panel).
}
\end{figure}



\begin{figure}
\includegraphics[width=84mm]{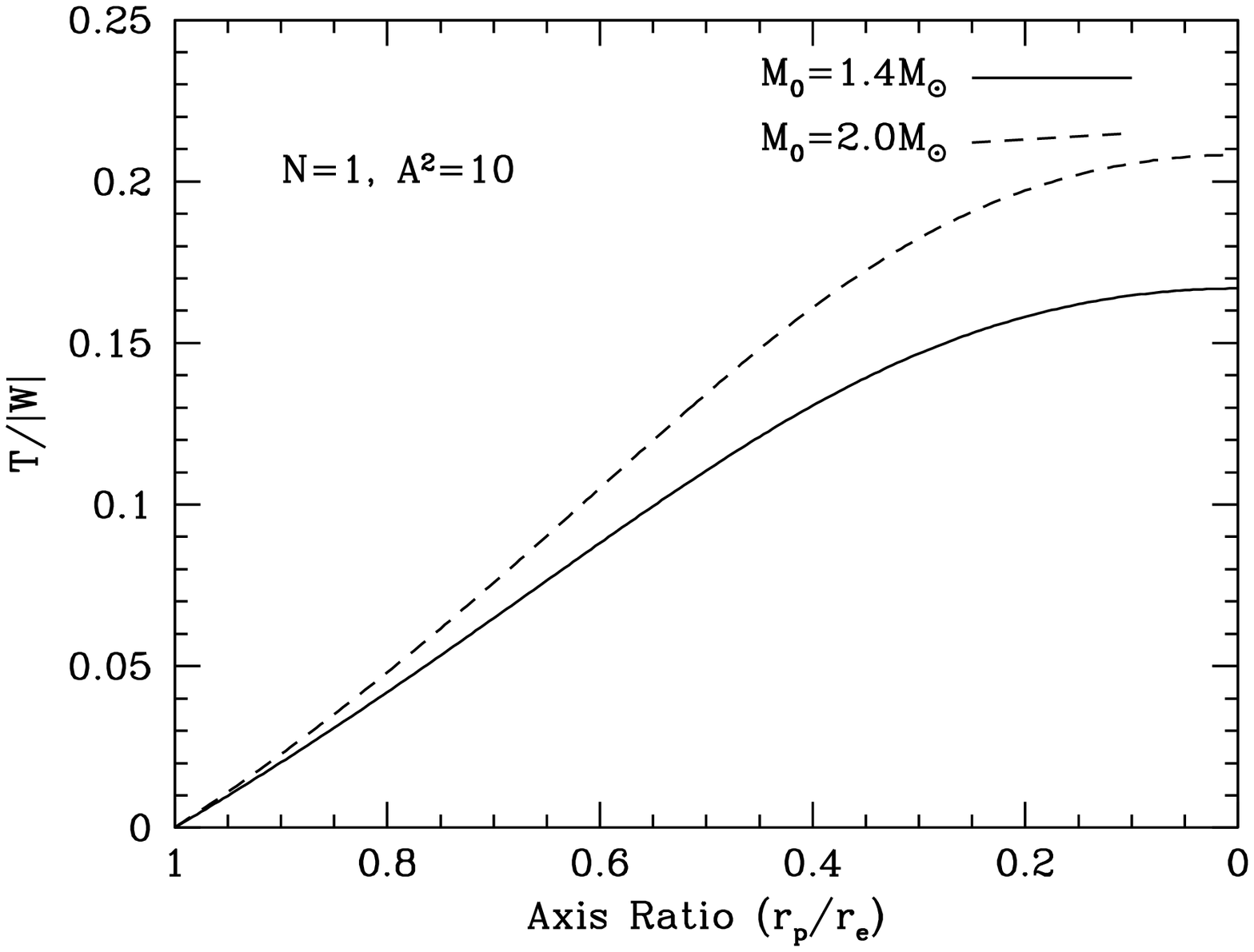}\vspace{-2cm}\\
\includegraphics[width=84mm]{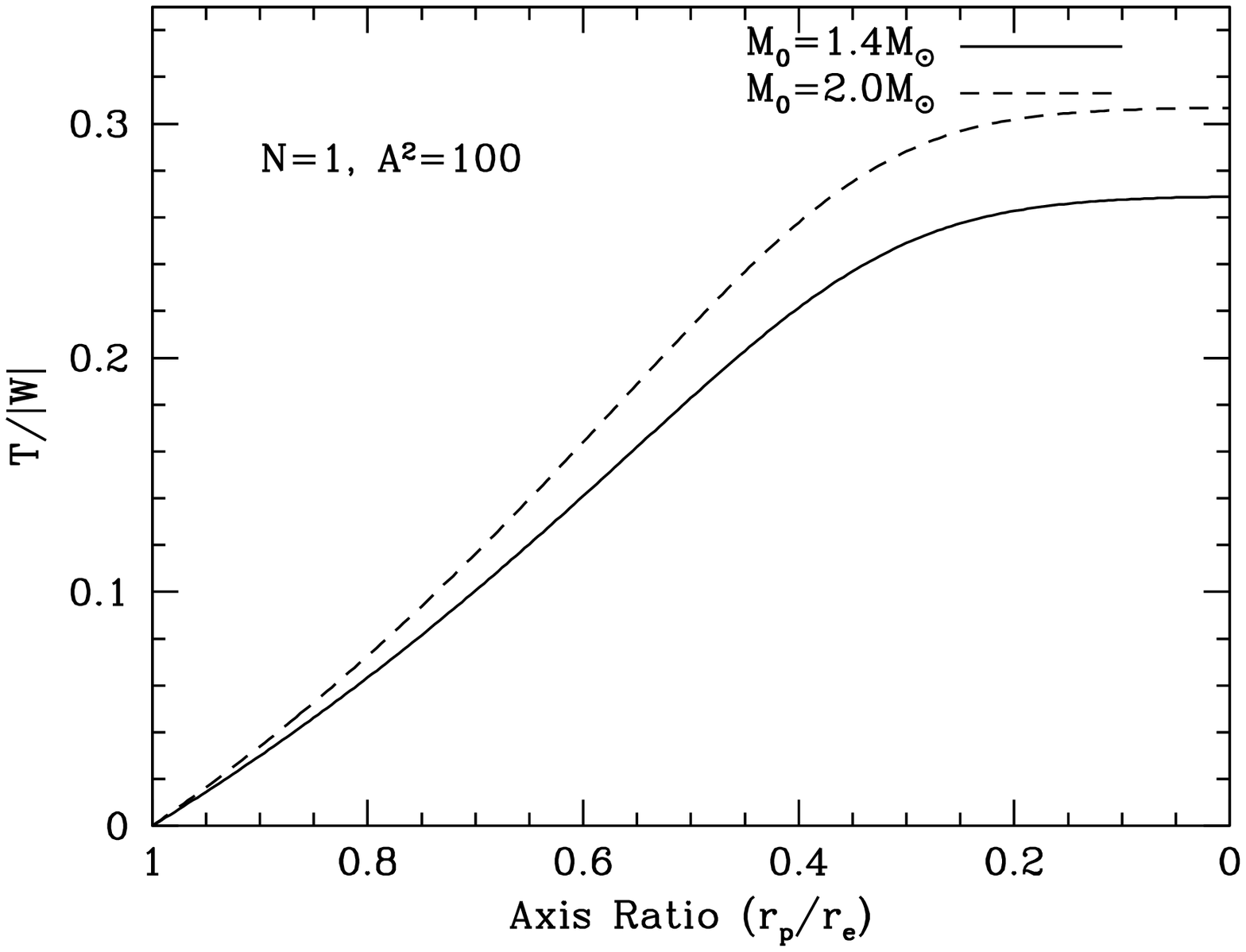}\vspace{-1cm}
\caption{The relationship between $T/|W|$ and axis ratio
for stars with $M_0=1.4M_{\odot}$ (solid) and $M_0=
2M_{\odot}$ (dotted) for rotation parameters of $A^2=10$ 
(top panel) and $A^2=100$ 
(bottom panel).
}
\end{figure}


Figure 12 shows $T/|W|$ as a function of the axis-ratio for the models 
shown in Figures 10 and 11. As we stated in the previous section, 
larger $A$ value makes the angular speed distribution nearly constant 
(i.e., uniform rotation).
Therefore, rotational energy ($T$) 
tends to become larger for larger $A$.
The most rapidly rotating models with $A^2=10$ have $T/|W|$=0.17 and
0.21 for the cases with baryon mass 1.4 and 2 $M_\odot$ 
while models with $A^2=100$ can have $T/|W|$=0.27 and 0.31 for 
$M=1.4$ and 2 $M_\odot$, respectively. 
Note, for differentially rotating stars, the onset
of the dynamical instabilty depends on the strength of the differential
rotation. For weak case, it turned out that $\beta_d 
\sim 0.27$ for Newtonian stars, just similar to the uniform rotation 
(see, for example, \cite{ll01}). 
However, many stuides have found that it could be 
down to $\beta_d = 0.2$ \citep{ke03}, $\beta_d = 0.14$ \citep{cen01, 
toh90,pdd96,ll01, liu02},
and even down to 0.04 \citep{ske02, ske03}.
Recent full GR simulation has found $\beta_d$ = 0.254 for weak
case \citep{bai07}.

\section{Summary and Discussion}
In order to find equilibrium solutions of rapidly rotating compact 
stars which have relativistic motions or relativistic equations of 
states, we have proposed an approximate method which is appropriate 
if the gravitational field is not very strong.
Assuming the weak gravity, the spacetime and the hydrostatic equations 
are derived by only considering the Newtonian gravitational potential.
However, in order to accommodate the relativistic effects, 
we have adopted the active mass density as a source for the 
gravitational potential. The active mass density takes into account 
all the forms of energy density.
The numerical calculation has been carried out by following the 
Hachisu's SCF method.
We have obtained the equilibrium solutions for a wide range of 
parameters and topological shapes such as spheroid, toroid, and 
quasi-toroid. Only the polytropic equation of state is considered 
for simplicity.
The inclusion of special relativistic effects could 
significantly improve the accuracy of the solutions if the rotational
speed is substantial compared to the purely Newtonian hydrodynamical
approach.  
We have shown that the solutions can be further improved by 
taking into account the the contribution from the active mass density
for the computation of the gravitational field.
Even for a mildly relativistic case ($R \sim 0.1$), we found that the
active mass plays important role. 

We have concentrated only on the equilibrium models in this paper.
The studies on the stability of the relativistic stars and their 
physical sequences should be followed by integrating dynamical
equations. In the subsequent works, we will provide the
full set of dynamical equations under the same assumptions (i.e.,
Newtonian gravity with active mass and special relativistic 
hydrodynamics) and carry out the dynamical evolution studies of
rapidly rotating compact stars.


\vskip 3mm
This work was supported by the Korean Research Foundation Grant No. 2006-341-C00018. 
JK acknowledges the support BK21 program to SNU.


\end{document}